\documentclass[twocolumn,aps,superscriptaddress]{revtex4-2}

\makeatletter

\newcommand{\Rmnum}[1]{\expandafter\@slowromancap\romannumeral #1@}
\makeatother

\usepackage{graphicx}% Include figure files
\usepackage{epstopdf}
\usepackage{epstopdf}
\usepackage{dcolumn}% Align table columns on decimal point
\usepackage{bm}% bold math

\usepackage[utf8]{inputenc}
\usepackage[T1]{fontenc}
\usepackage{booktabs, array, mathptmx, float, tabularx, booktabs, lipsum, amsmath,multirow}
\usepackage{siunitx, xcolor}
\usepackage[version=4]{mhchem}
\graphicspath{{figs/}{figsgaoerb/}} % 读取图片从figs/文件夹中读取，而不是当前文件夹
\usepackage[colorlinks,linkcolor=blue,anchorcolor=blue,citecolor=blue]{hyperref}
\begin{document}

\bibliographystyle{plain}

%\preprint{}

\title{Goos-Hänchen Shift and Photonic Spin Hall Effect in Semi-Dirac Material Heterostructures}


\author{Hong-Liang Cheng}
\affiliation{College of Physics, Nanjing University of Aeronautics and Astronautics, Nanjing 210016, People’s Republic of China}

\author{Yong-Mei Zhang}
\thanks{zymzym@nuaa.edu.cn}
%\email{zymzym@nuaa.edu.cn}
\affiliation{College of Physics, Nanjing University of Aeronautics and Astronautics, Nanjing 210016, People’s Republic of China}
\affiliation{Key Laboratory of Aerospace Information Materials and Physics (NUAA), MIIT, Nanjing 211106, People’s Republic of China}

%\thanks{}


\date{\today}

\begin{abstract}
We investigate the photonic spin Hall effect (PSHE) and the Goos-Hänchen shift (GH shift) in semi-Dirac materials. Through theoretical modeling, we demonstrate that the anisotropic dielectric function in semi-Dirac materials play a critical role in determining the magnitude and polarity of these optical displacements. Furthermore, by utilizing the unidirectional drift of massless Dirac electrons in Semi-Dirac materials, we systematically reveal how the drift velocity and direction modulate the behavior of optical displacements. The results indicate that semi-Dirac materials provide a versatile platform for controlling spin-dependent photonic phenomena with their material anisotropy and carrier transport. This work opens a new avenue for designing advanced photonic devices with tunable optical responses, particularly with significant application potential in quantum information processing and topological photonics.    % insert abstract here
\end{abstract}

%\keywords{semi-Dirac; photonic spin Hall effect; Goos-Hänchen shift; Fizeau drag.}


\maketitle
\section{\label{sec:level1}Introduction}
In recent years, the photonic spin Hall effect (PSHE) \cite{1,2,3} and the Goos-Hänchen (GH) shift \cite{4,xiang22,huang24} induced by light reflection at interfaces have garnered significant attention due to their potential applications in nanophotonics \cite{5}, topological photonics \cite{Da2023,Da2021}, optical device design\cite{Liu2024,DENG2024}, high-energy physics \cite{6}, and the study of properties in two-dimensional materials \cite{7,8,Muzamil2024,Shah:22,Jahani:23,SALIM2024}. The PSHE manifests a transverse spin-dependent splitting of linearly polarized light into left-circularly polarized (LCP) and right-circularly polarized (RCP) components, serving as an optical analogy to the electronic spin Hall effect \cite{Dong_2020}. In contrast, the GH shift describes the longitudinal spatial displacement of a light beam upon reflection. Both phenomena originate from the spatial gradient of the reflection phase at the interface and the coupling between light-matter interactions. However, traditional studies have primarily focused on isotropic media or metal interfaces \cite{Zhu2021,Xia2021,Zhang2020,Wan2020}, leaving a significant gap in the analysis of polarization-dependent responses in two-dimensional anisotropic materials and heterojunction systems.

Semi-Dirac materials\cite{Real2020,HUANG2023} have become a research hotspot due to their unique band structures: they exhibit linear dispersion (Dirac-like cone) \cite{Tabatabaei2022,DonísVela2022,Gao2023} along one direction and parabolic dispersion (conventional semiconductor-like) along the perpendicular direction \cite{9,10}. This anisotropic property endows semi-Dirac materials with both the massless carrier characteristics of Dirac materials and the tunability of traditional semiconductors. Semi-Dirac states have been realized in $TiO_{2}/VO_{2}$ heterojunctions \cite{11}, strained graphene \cite{12,Duan2017}, black phosphorus \cite{13}, and organic conductors \cite{14}. For instance, in strained graphene, semi-Dirac states can be induced through lattice deformation, resulting in a strongly anisotropic conductivity tensor \cite{15}, which provides a new degree of freedom for light-field manipulation. However, existing studies have predominantly focused on the electronic transport properties of semi-Dirac materials, while their photonic potential—especially the polarization-dependent responses of PSHE and GH shifts—remains largely unexplored.

Meanwhile, the Fizeau drag effect \cite{16,Banerjee2022}, a classical phenomenon of light speed modulation in moving media, has recently revealed new implications in two-dimensional material systems. Leveraging the high mobility of massless Dirac electrons and the slow plasmon propagation characteristics in graphene, experiments have demonstrated that the momentum coupling between light waves and drifting electrons can significantly alter the effective dielectric response \cite{17,18}. In semi-Dirac materials, massless electrons along the linear dispersion direction exhibit directional drift ($v_{D}$) under an electric field, providing a unique platform for actively controlling optical displacements. One research report demonstrated the achievement of non-reciprocal optical modulation effects through carrier drift in current-driven graphene heterostructures \cite{BARVESTANI2024}. However, existing studies have yet to systematically investigate the impact of electron drift in semi-Dirac materials on PSHE and GH shifts, particularly in terms of the coupling between anisotropic conductivity and nonreciprocal optical responses, where a quantitative model remains lacking.

To address these challenges, this paper proposes an optical modulation system based on a carrier-driven semi-Dirac material-dielectric heterojunction and systematically investigates the characteristics of PSHE and GH shifts induced by S/P-polarized light incidence, revealing the regulatory role of the Fizeau drag effect.

The structure of this paper is as follows: Section \Rmnum{2} establishes the theoretical model, derives the Fresnel coefficients for the semi-Dirac material-dielectric heterojunction, formulates the GH shift for S/P-polarized light and the PSHE shift for P-polarized light, and incorporates the influence of the Fizeau Drag Effect on optical displacement. Section \Rmnum{3} employs numerical calculations to reveal the modulation of displacement by electron drift velocity, incident angle, and wavelength, comparing the nonreciprocal responses of reverse ($v_{D}<0$) and forward ($v_{D}>0$) drift modes. Section \Rmnum{4} summarizes the key findings and explores potential applications in topological photonic devices and quantum information processing.\\%
\section{\label{sec:level1}	Model and Theory}
\begin{figure*}[t!]
	\centering
	\includegraphics[width=0.8\linewidth]{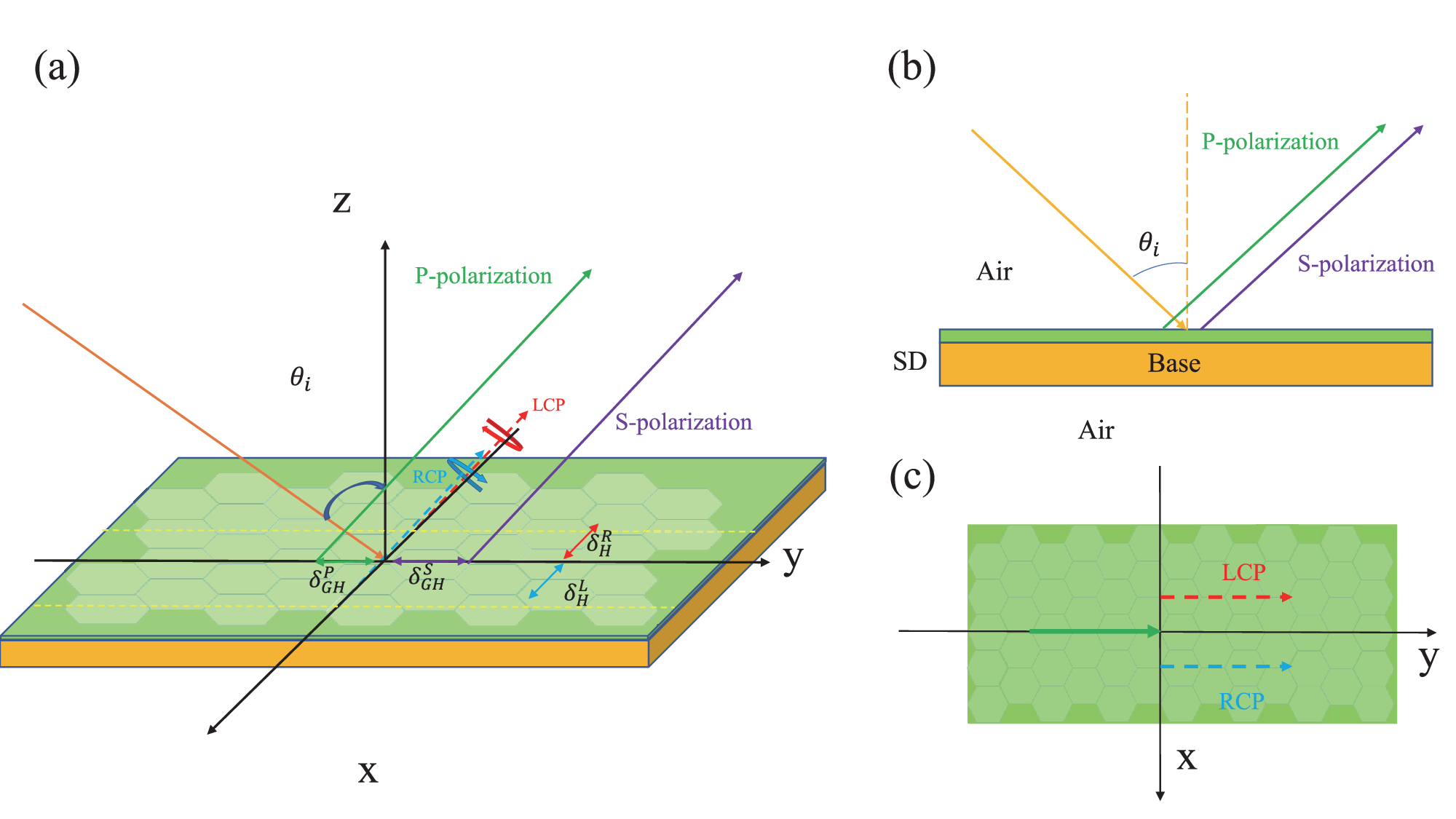}
	\caption{
		Schematic illustration for studying the GH shift and PSHE at the surface of a semi-Dirac material. 
		The semi-Dirac material, with a thickness of $d_2$, is deposited on a substrate 
		with a thickness of $d_3$. $\delta_{\mathrm{GH}}^{(S,P)}$ represents the longitudinal 
		displacement (GH shift) for $S$- and $P$-polarized light, while $\delta_{H}^{(L,R)}$ 
		represents the transverse displacement (PSHE shift) of the left-circularly polarized 
		(LCP) and right-circularly polarized (RCP) components split from horizontally polarized light.
	}
	\label{fig_1}
\end{figure*}

In this work, we calculate the GH effect and PSHE effect for a system composed of air, a semi-Dirac material, and a substrate (FIG.~\ref{fig_1}). The relative permittivities of the media above and below the semi-Dirac material are $\varepsilon_1=1$ and $\varepsilon_3=2.25$, respectively, while the region below the substrate is also air with $\varepsilon_4=\varepsilon_1$. The substrate thickness is $\SI{2.5}{\micro\meter}$.

According to the Fresnel reflection formulas, the reflection coefficients $r_{A,B}^{S}$ and $r_{A,B}^{P}$ represent the reflection coefficients of S- and P-polarized light at the interface between media $A$ and $B$, respectively:
\begin{align}
	r_{A,B}^{S} &= \frac{n_B \cos\theta_B - n_A \cos\theta_A}{n_B \cos\theta_B + n_A \cos\theta_A} \\
	r_{A,B}^{P} &= \frac{n_B \cos\theta_A - n_A \cos\theta_B}{n_B \cos\theta_A + n_A \cos\theta_B}
\end{align}
Here, $A,B=1,2,3,4$ represent the upper air, semi-Dirac material, substrate, and lower air, respectively. $\theta_i = \theta_1$ is the angle of incidence, while $\theta_{2,3,4}$ are the refraction angles in media 2, 3, and 4, determined by Snell's law across the two media in the structure. $n_A$ and $n_B$ denote the refractive indices of media $A$ and $B$, respectively. Taking into account the multiple reflection effects in media 2, 3, and 4, the expressions are given by \cite{19}:
\begin{align}
	r_{234}^{S,P} &= \frac{r_{23}^{S,P} + r_{34}^{S,P} \exp(2i k_{3z} d_3)}{1 + r_{23}^{S,P} r_{34}^{S,P} \exp(2i k_{3z} d_3)} 
\end{align}
here, $r_{A,B}^{(S,P)}$ represents the reflectance of S- and P-polarized light at the interface between media $A$ and $B$, $d_3$ is the thickness of medium 3. 
$k_{iz} = \sqrt{k_i^2 - k_{iy}^2}$ and $k_i$ is the wavenumber in medium $i$. The total reflection coefficients for S- and P-polarized light in the system are given by:
\begin{align}
	r^{S,P} &= \frac{r_{12}^{S,P} + r_{234}^{S,P} \exp(2i k_{2z} d_2)}{1 + r_{12}^{S,P} r_{234}^{S,P} \exp(2i k_{2z} d_2)}
\end{align}

The GH shift is a spatial displacement caused by the phase change of reflected light. The spatial GH shift is expressed as \cite{20}:
\begin{equation}\label{eq5}
	\delta_{GH} = \frac{\lambda}{2\pi} \mathrm{Im}(D_r)
\end{equation}
\begin{equation}
	D_r = \frac{\partial \ln r}{\partial \theta_i}= \frac{1}{R} \frac{\partial R}{\partial \theta_i} + i \frac{\partial \Phi_r}{\partial \theta_i}
\end{equation}
where $R = |r|$ is the amplitude of the reflection coefficient, $\Phi_r$ is the phase of the reflection coefficient amplitude.

For horizontally (H) polarized (P-polarized) incident light, the optical displacement of the reflected beam $\delta_H^{(L,R)}$ is determined by the polarization type of the incident light, the reflection coefficient, and the angle of incidence $\theta_i$. Assuming the full width at half maximum of the reflection coefficient is not narrow, $\delta_H^{(L,R)}$ can be expressed as \cite{21}:
\begin{equation}\label{eq7}
	\delta_H^{(L,R)} = \mp\left(1 + \frac{R_S \cos(\Phi_S - \Phi_P)}{R_P}\right) \frac{\cot\theta_i}{k_0}
\end{equation}
where $k_0$ is the wave vector in free space. The transverse displacement for vertically polarized (S-polarized) incident light is too small and will not be discussed here.

The expression for the dielectric function $\epsilon_{\mathrm{SD}}(\omega)$ of a single-layer semi-Dirac material is given by \cite{22,Yang2025}:
\begin{equation}
	\epsilon_{\mathrm{SD}}(\omega) = \frac{\epsilon_1 + \epsilon_3}{2} + \frac{i\sigma}{\omega\epsilon_0 t_{\mathrm{SD}}}
\end{equation}
where $t_{\mathrm{SD}}$ is the thickness of a single-layer semi-Dirac material ($t_{\mathrm{SD}} = 5a_0$). Thickness converts the three-dimensional material's dielectric constant to two-dimensional. Here, $\epsilon_1$ and $\epsilon_3$ are the relative permittivities of medium 1 and medium 3, respectively, $\omega$ is the angular frequency. $\sigma$ is the conductivity, determined by the material's band structrure.  Under the tight-binding approximation, the Hamiltonian of the semi-Dirac material can be written as \cite{15}:
\begin{widetext}
\begin{equation}
	H_0 = \sum_{i,j}\left[t_2 c_i^\dagger(R_i) c_j(R_i+\delta_1) + t_1 c_i^\dagger(R_i) c_j(R_i+\delta_2) + t_1 c_i^\dagger(R_i) c_j(R_i+\delta_3)\right] + \mathrm{h.c.}
\end{equation}
\end{widetext}

Here, the operator $c_i^\dagger(R_i)$ represents the creation of an electron at the lattice site $R_i$, while $c_j(R_j)$ represents the annihilation of an electron at the lattice site $R_j$. $t_1$, $t_2$ are the hopping parameters between the nearest neighbors. In semi-Dirac system, we have $t_2 = 2t_1$. The Hamiltonian in reciprocal space is written as:
\begin{equation}
	H_0 = \sum_{\mathbf{k}} \begin{pmatrix} c_A^\dagger & c_B^\dagger \end{pmatrix}
	\begin{pmatrix} 
		0 & H_{12} \\
		H_{21} & 0 
	\end{pmatrix}
	\begin{pmatrix} c_A \\ c_B \end{pmatrix}
\end{equation}
with
\begin{align}
	H_{12} &= -(t_2 e^{-i\mathbf{k}\cdot\delta_1} + t_1 e^{-i\mathbf{k}\cdot\delta_2} + t_1 e^{-i\mathbf{k}\cdot\delta_3}) \\
	H_{21} &= H_{12}^*
\end{align}
where $\delta_1$, $\delta_2$, $\delta_3$ represent the relative position vectors of the three nearest neighbor lattice points.

Combining the Hamiltonian, the conductivity formula for semi-Dirac materials at temperature $T$ is given by the Kubo formula \cite{23,HungNguyen2017,Rashidian2014}:
\begin{align}
	\sigma_{\alpha\beta}^{\mathrm{inter}}(\omega) &= \frac{ie^2\hbar}{\omega S} \sum_{\mathbf{k},n,m} \frac{(V_{\mathbf{k}}^\alpha V_{\mathbf{k}}^\beta)(f_{\mathbf{k},m} - f_{\mathbf{k},n})}{E_{m}(\mathbf{k}) - E_{n}(\mathbf{k}) - (\hbar\omega + i\eta_1)} \\
	\sigma_{\alpha\beta}^{\mathrm{intra}}(\omega) &= -\frac{ie^2\hbar}{S(\hbar\omega + i\eta_2)} \sum_{\mathbf{k},n,m} (V_{\mathbf{k}}^\alpha V_{\mathbf{k}}^\beta) \left(-\frac{\partial f(\mathbf{k})}{\partial E_{j}}\right) \\
	\sigma &= \sigma^{\mathrm{inter}} + \sigma^{\mathrm{intra}}
\end{align}
where $(V_{\mathbf{k}}^\alpha V_{\mathbf{k}}^\beta) = \langle \mathbf{k},n|V^\alpha(\mathbf{k})|\mathbf{k},m \rangle \langle \mathbf{k},m|V^\beta(\mathbf{k})|\mathbf{k},n \rangle$, $V^\alpha(\mathbf{k}) = \frac{1}{\hbar} \frac{\partial H(\mathbf{k})}{\partial k_\alpha}$ is velocity of energy band electrons, $\alpha, \beta \in \{x,y,z\}$. $f_n(\mathbf{k}) = \left[e^{E_{n}(\mathbf{k})/(k_B T)} + 1\right]^{-1}$ represents the Fermi distribution function for the $n$-th energy band. $S$ denotes the area of the semi-Dirac system, $E_{n}(\mathbf{k})$ is the eigenenergy, and $\eta$ is a damping factor ($\eta_1 = 0.01\ \mathrm{eV}$, $\eta_2 = 2.6\times10^{-3}\ \mathrm{eV}$).

In semi-Dirac materials, the physical quantities of massless Dirac electrons under the influence of a magnetic field or other external forces are adjusted through the Lorentz transformation \cite{24,Blevins2024}:
\begin{align}
	\omega^m &= \gamma(\omega - v_D k_2) \\
	k_{\mathrm{SD}}^m &= \gamma\left(k_2 - \frac{v_D}{v_F^2} \omega\right)
\end{align}
The relativistic dynamics of drifting Dirac electrons in semi-Dirac materials is governed by the Lorentz factor $\gamma = 1\big/\sqrt{1 - v_D^2/v_F^2}$ \cite{17}, where m represents ($+$) or ($-$) (indicating the electron drift direction is the same as or opposite to the light incident direction) .  $v_D$ denotes the electron drift velocity, $v_F$ the Fermi velocity. 
When considering the Fizeau drag effect caused by the drift of massless Dirac electrons, the phase velocity of the wave becomes: $v^m = v \pm v_D F$, with the drag coefficient $F = n_{g2}/n_{\mathrm{SD}} - 1/n_{\mathrm{SD}}^2$ \cite{24} derived from the group velocity refractive index $n_{g2} \equiv n_{\mathrm{SD}} + \omega(\partial n_{\mathrm{SD}}/\partial\omega)$. By substituting the results of the Lorentz transformation into the phase velocity, we obtain the wave vectors $k_{\mathrm{SD}}^+$ and $k_{\mathrm{SD}}^-$ for co-propagating and counter-propagating waves, respectively, expressed as:
\begin{widetext}
\begin{align}
	k_{\mathrm{SD}}^+ &= \frac{
		- (v_D \gamma \omega F + \gamma v_F^2 k_{2})v_D 
		\pm \gamma v_D \sqrt{
			k_{2}^2 v_F^4 
			- 2 v_F^2 v_D \omega k_{2} F 
			+ 4 v_F^2 \omega^2 F 
			+ v_D^2 \omega^2 F^2
		}
	}{
		2 v_F^2 v_D F
	} \\
	k_{\mathrm{SD}}^- &= \frac{
		(-v_D \gamma \omega F + \gamma v_F^2 k_{2})v_D 
		\pm \gamma v_D \sqrt{
			k_{2}^2 v_F^4 
			+ 2 v_F^2 v_D \omega k_{2} F 
			- 4 v_F^2 \omega^2 F 
			+ v_D^2 \omega^2 F^2
		}
	}{
		2 v_F^2 v_D F
	}
\end{align}
\end{widetext}

Through the above theoretical analysis, we systematically derive the reflection effects of light on a single-layer semi-Dirac material, including the Goos-Hänchen shifts for $S$- and $P$-polarized light and the transverse shifts caused by the photonic spin Hall effect for $P$-polarized light. Eq. ( \ref{eq5} ) and Eq. ( \ref{eq7} ) provide the theoretical foundation for calculating the spatial displacements of the reflected beam, particularly taking into account the effects of spin coupling and material properties.

\section{\label{sec:level1}Results and Discussion}
\subsection{\label{sec:level2}PSHE and GH Shift under Unperturbed Conditions}
\begin{figure}
	\centering
	\includegraphics[width=0.8\linewidth]{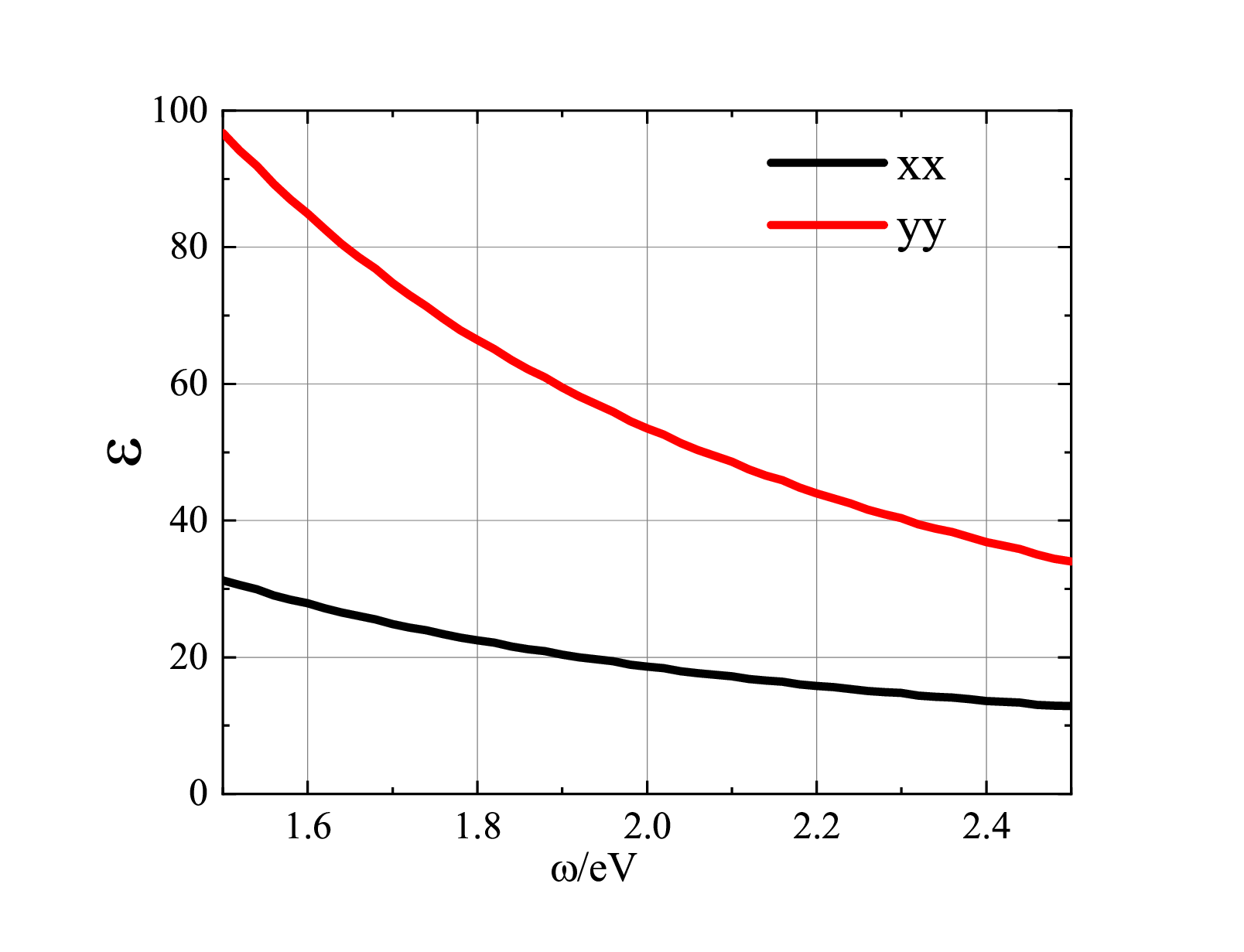}
	\caption{
		The effective dielectric constant of the semi-Dirac material in this system: (a) $xx$ component and (b) $yy$ component. $T=300K$.
	}
	\label{fig_2}
\end{figure}

We first focus on the anisotropy of semi-Dirac materials to investigate the effects of semi-Dirac materials on the reflect light. Due to the band structure of semi-Dirac materials exhibiting a nonlinear relationship (free electron gas) in one direction and a linear relationship (massless Dirac electrons) in the perpendicular direction, there is a significant difference between the $xx$ and $yy$ components of their dielectric constant, as shown in FIG.~\ref{fig_2}. The refractive index influenced by the dielectric constant also demonstrates anisotropy.

\begin{figure}
	\centering
	\includegraphics[width=1\linewidth]{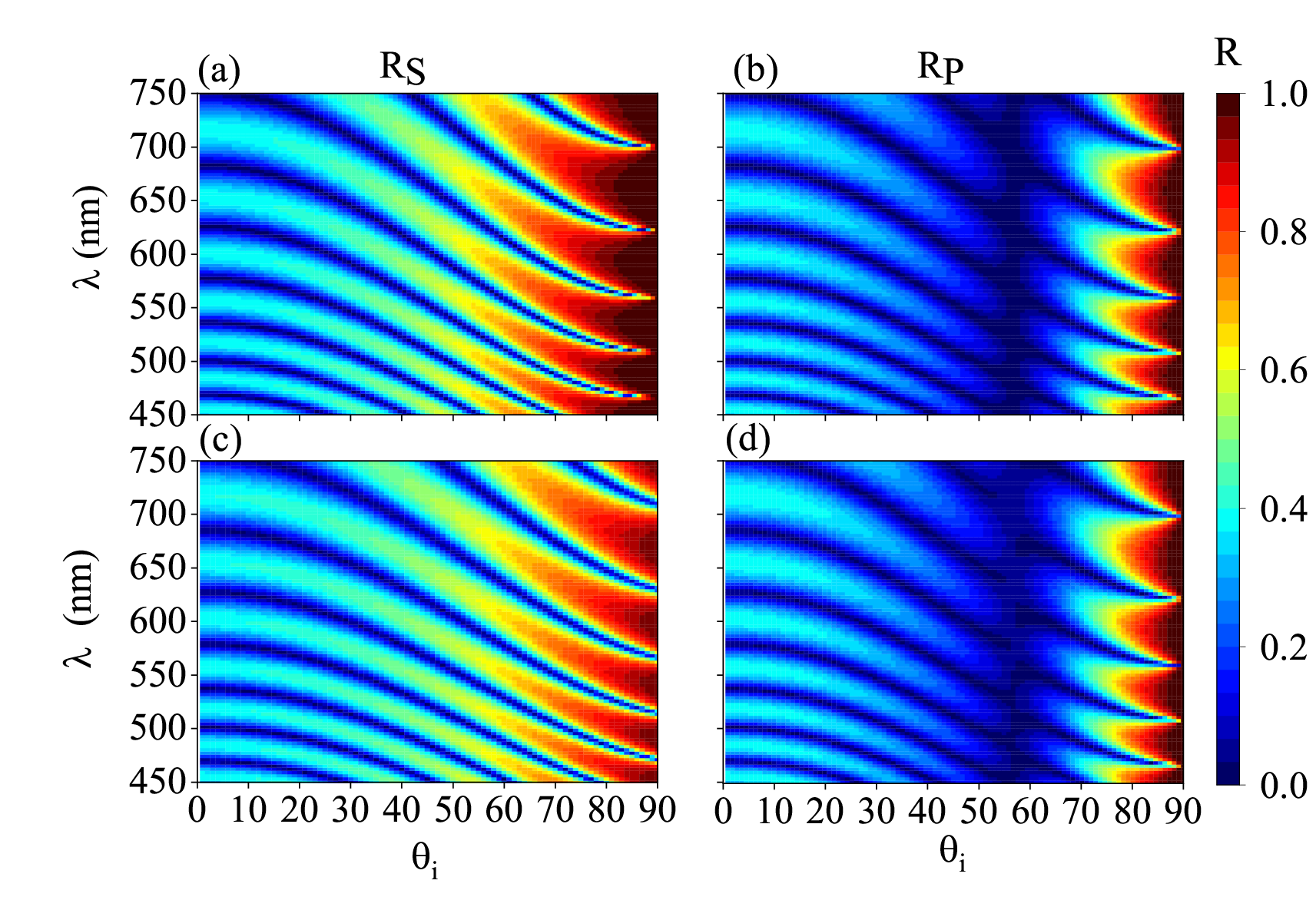}
	\caption{
	The reflection coefficients $R = |r|$ for S- and P-polarized light as functions of the incident angle $\theta_i$ and the incident light wavelength $\lambda$: (a)(b) $\varepsilon = \varepsilon_{xx}$, (c)(d) $\varepsilon = \varepsilon_{yy}$.	
	}
	\label{fig_3}
\end{figure}

To examine how incident light wavelength $\lambda$ and the incident angle $\theta_i$ influence the optical displacement of the system, FIG.~\ref{fig_3} presents the reflection coefficients $R_S$ and $R_P$ for $S$- and $P$-polarized light as functions of $\theta_i$ and $\lambda$ under the $xx$ and $yy$ components of the dielectric constant of the semi-Dirac material. Specifically, (a) and (b) depict the case for $\varepsilon = \varepsilon_{xx}$, while (c) and (d) depict the case for $\varepsilon = \varepsilon_{yy}$. It can be observed that as $\lambda$ and $\theta_i$ vary, both $R_S$ and $R_P$ exhibit an approximately periodic trend. The reflection coefficient $R_P$ approaches zero near 57°, consistent with Brewster's law. This indicates that when light is incident at the Brewster angle, the reflection of $P$-polarized light is suppressed. Both $R_S$ and $R_P$ have lower reflection coefficients in the low incident angle region compared to the high incident angle region. A pairwise analysis of the Fig.~\ref{fig_3} panels—comparing (a) with (c), and (b) with (d)—reveals a systematic shift of the reflection coefficients for both S- and P-polarized light toward larger incidence angles. This behavior can be attributed to dielectric anisotropy, specifically due to $\varepsilon_{yy} > \varepsilon_{xx}$.
\begin{figure}
	\centering
	\includegraphics[width=1\linewidth]{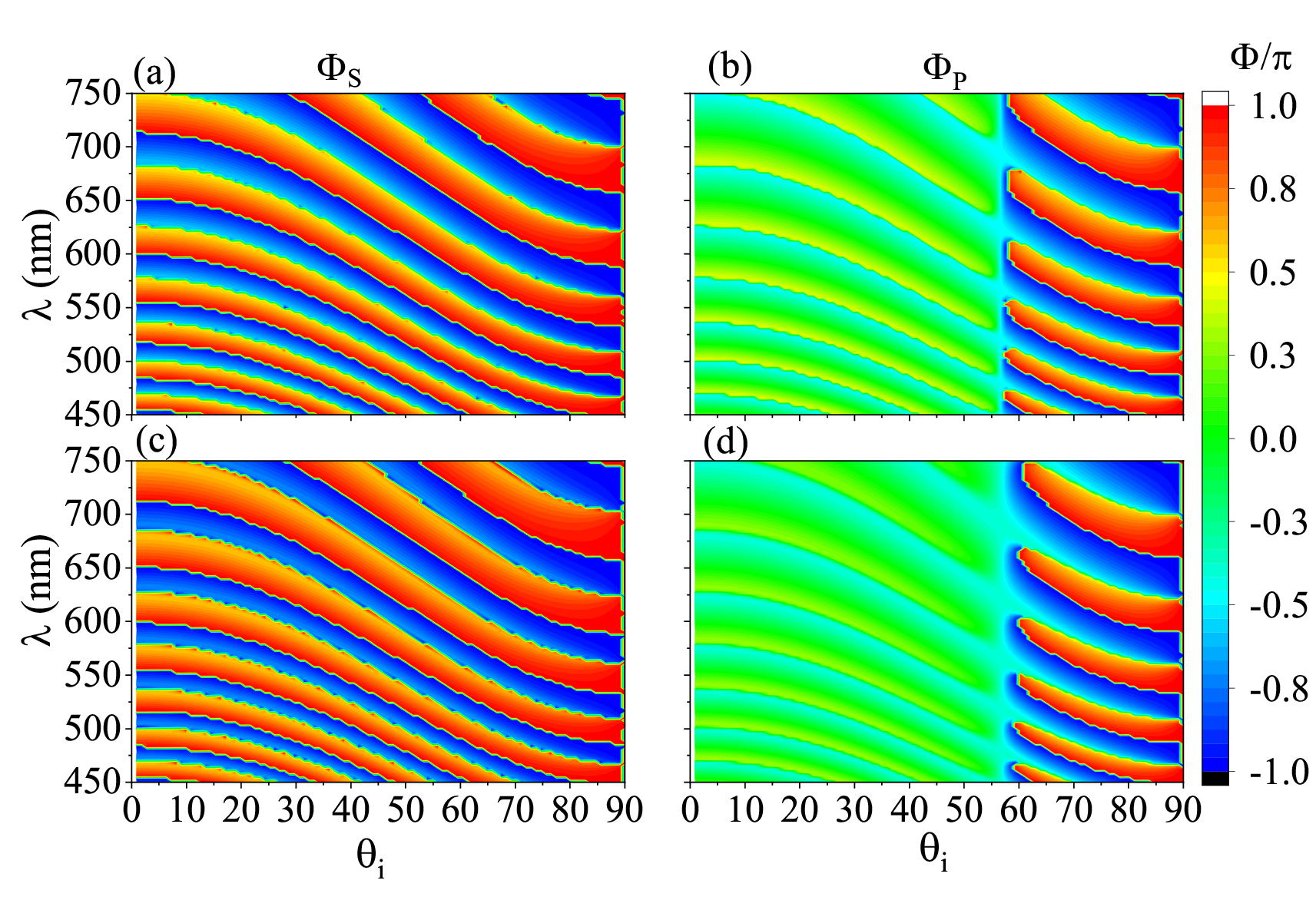}
	\caption{
		The reflection phase diagrams for S- and P-polarized light as functions of the incident angle $\theta_i$ and the incident light wavelength $\lambda$: (a)(b) $\varepsilon = \varepsilon_{xx}$, (c)(d) $\varepsilon = \varepsilon_{yy}$.	
	}
	\label{fig_4}
\end{figure}

Changes in the incident angle $\theta_i$ and the incident light wavelength $\lambda$ not only alter the reflection coefficients but also affect the phase of the reflection coefficients, as shown in FIG.~\ref{fig_4}. For $S$-polarized light, as $\theta_i$ increases from $0$ to $\pi/2$, multiple phase transition points appear, and the number of these points decreases as $\lambda$ increases. For $P$-polarized light, the phase transition points always occur in regions beyond the Brewster angle. The magnitude of $\varepsilon$ has minimal impact on the overall phase diagram.

\begin{figure}
	\centering
	\includegraphics[width=1\linewidth]{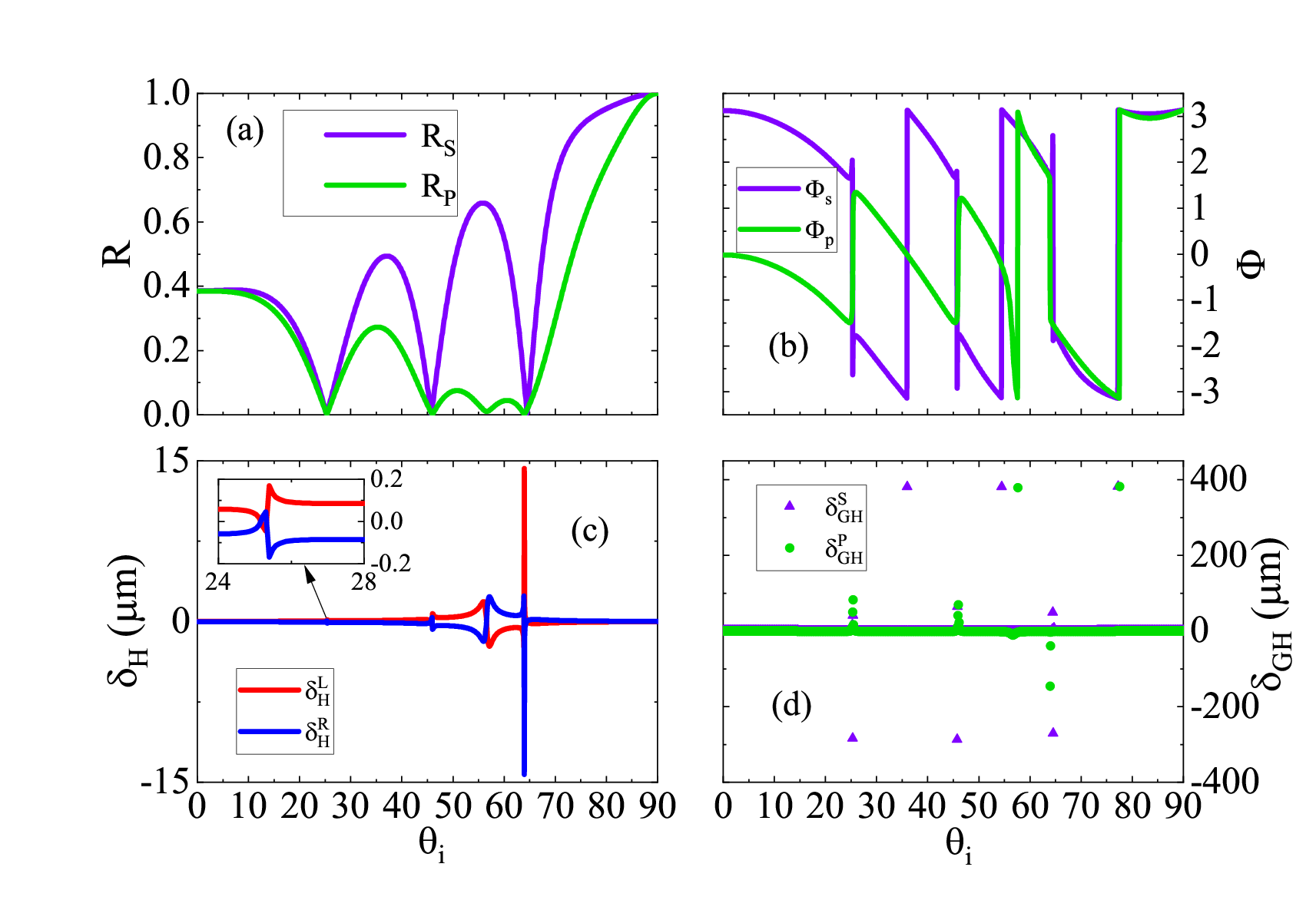}
	\caption{
		Dependence of (a) reflection coefficient, (b) phase of the reflection coefficient for S- and P-polarized light, (c) PSHE shift for P-polarized light, (d) GH shift for $S$- and $P$-polarized light on the incident angle $\theta_i$. The purple and green solid lines correspond to $S$- and $P$-polarized light, while the red and blue solid lines correspond to left- and right-circularly polarized light. The parameters are set as $\lambda = 600$ nm, $\varepsilon_3 = 2.25$, $v_D = 0$, and $\varepsilon = \varepsilon_{xx}$.	
	}
	\label{fig_5}
\end{figure}

We choose the case of $\lambda=600$ nm. FIG.~\ref{fig_5}(a)(b) and FIG.~\ref{fig_6}(a)(b) present the reflectance coefficients and phases for $S$- and $P$-polarized light in a system based on $\varepsilon = \varepsilon_{yy}$ and $\varepsilon = \varepsilon_{yy}$. It can be observed that the system exhibits reflectance minima and phase transition points at certain angles. Moreover, the positions of the reflectance zeros and phase transition points do not completely coincide, which is consistent with the results in FIG.~\ref{fig_3} and FIG.~\ref{fig_4}.

From FIG.~\ref{fig_5}(c) and FIG.~\ref{fig_6}(c), it can be seen that a significant PSHE shift appears near the reflectance minima for $S$- and $P$-polarized light, and an enormous PSHE shift occurs near the zero reflectance of $P$-polarized light.The peak values of the photonic spin Hall effect (PSHE) exhibit marked differences under two distinct dielectric tensor components, with significantly larger PSHE displacements observed in systems possessing smaller dielectric components. This phenomenon aligns with the theoretical framework established by Cheng et al.\cite{Cheng2022}. Meanwhile, at the phase transition points of the reflection coefficient for $S$- and $P$-polarized light, a large longitudinal displacement (GH shift) is observed.
The giant GH shift reaches 400 $\mu m$, representing approximately two orders of magnitude enhancement compared to Weyl semimetal structures \cite{Wu2021}.
The high dielectric constant of the $yy$ component in semi-Dirac materials leads to an increase in the minimum values of the reflection coefficients for both $S$- and $P$-polarized light. Under the influence of the $yy$ component, the PSHE shift of the system is significantly reduced compared to that under the $xx$ component.

However, the large GH shifts remain consistent under both dielectric components.
\subsection{\label{sec:level2}PSHE and GH shift with Reverse Fizeau drag}
\begin{figure}
	\centering
	\includegraphics[width=1\linewidth]{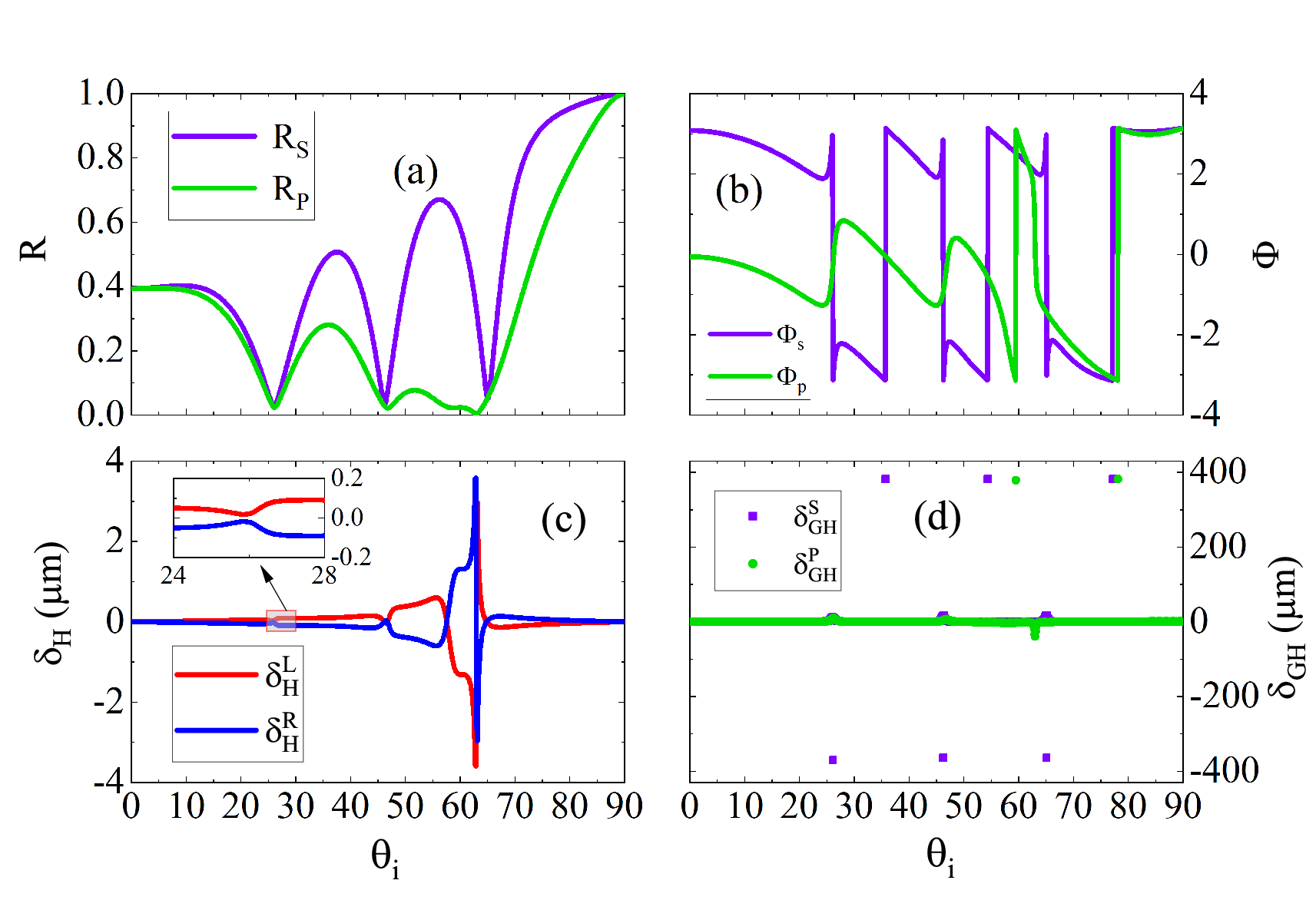}
	\caption{
		Dependence of (a) reflection coefficient, (b) phase of the reflection coefficient for $S$- and $P$-polarized light, (c) PSHE shift for $P$-polarized light, (d) GH shift for $S$- and $P$-polarized light on the incident angle $\theta_i$. The purple and green solid lines correspond to $S$- and $P$-polarized light, while the red and blue solid lines correspond to left- and right-circularly polarized light. The parameters are set as $\lambda=600$ nm, $\varepsilon_3=2.25$, $v_D=0$, $\varepsilon=\varepsilon_{yy}$.	
	}
	\label{fig_6}
\end{figure}

\begin{figure}
	\centering
	\includegraphics[width=1\linewidth]{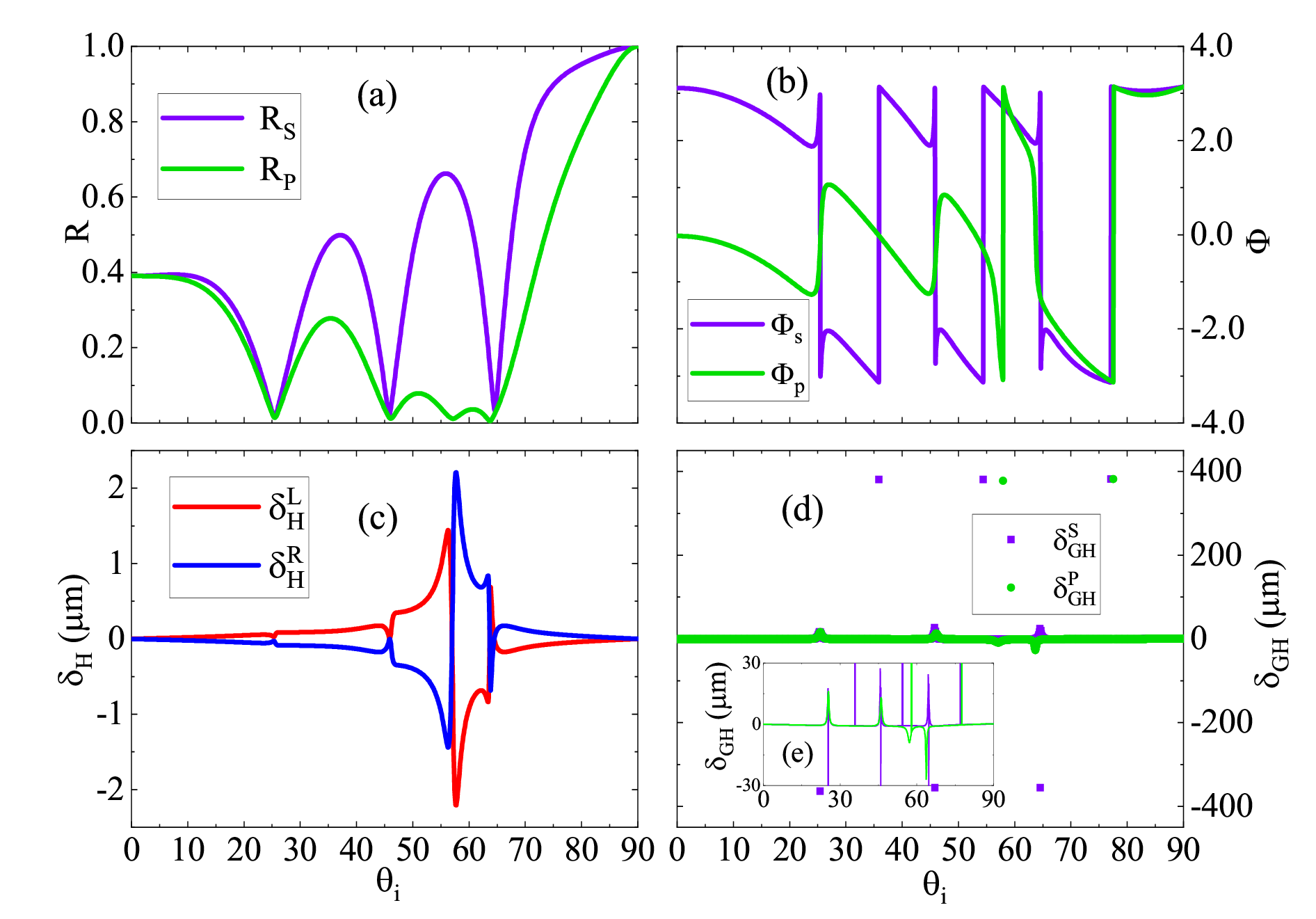}
	\caption{
		Dependence of (a) Reflection coefficients for $S$- and $P$-polarized light, (b) PSHE shift for $P$-polarized light, (c) GH shift for $S$-polarized light, and (d) GH shift for $P$-polarized light on the incident angle $\theta_i$ with reverse Fizeau drag ($v_D= -0.35v_F$). The purple and green solid lines correspond to $S$- and $P$-polarized light, while the red and blue solid lines correspond to left- and right-circularly polarized light. The parameters are set as $\lambda=600$ nm, $\varepsilon_3=2.25$, $\varepsilon=\varepsilon_{yy}$.		
	}
	\label{fig_7}
\end{figure}

Since the directional drift of massless Dirac electrons can induce Fizeau drag effects that modify light propagation in the medium, we exclusively focus on the dielectric component $\varepsilon_{yy}$ in semi-Dirac materials.We present numerical results for the PSHE shift and GH shift of the reflected beam in the model under the Fizeau effect. As shown in FIG.~\ref{fig_7}, we first examine the case where the drift direction of Dirac electrons in the semi-Dirac material is opposite to the direction of light incidence (reverse drift). The electron drift velocity is set to $v_D = -0.35 v_F$. Compared to FIG.~\ref{fig_6}, we observe a significant decrease in the minimum values of the reflection coefficients for both $S$- and $P$-polarized light. Since the reflection coefficient for $P$-polarized light reaches a minimum near 57 degrees, the PSHE shift shows a peak increase around this angle, but its magnitude is only half of the maximum peak in FIG.~\ref{fig_6}(c). Additionally, the phases of the system's reflection coefficients are also affected. The phase jump point for $S$-polarized light shifts slightly to the left, leading to a corresponding small leftward shift in the positions where large GH shifts occur for the reflected $S$-polarized light.

FIG.~\ref{fig_8} demonstrates the influence of reverse-drifted Dirac electrons on the system under different drift velocities $v_D$. In FIG.~\ref{fig_8}(a), the solid lines represent the reflection coefficients for $S$-polarized light, while the dashed lines correspond to $P$-polarized light. The drift velocities are set to $v_D=-0.001,-0.35,-0.7v_F$. The electron drift velocity can be tuned by applying a direct current via a power source [17]. As $v_D$ increases, the reflection coefficients for both $S$- and $P$-polarized light remain nearly unaffected. However, the main peak amplitude of the PSHE shift decreases with higher $v_D$ (only the PSHE shift for left circularly polarized light is plotted, as the shifts for left- and right-circularly polarized light only differ in sign), as shown in FIG.~\ref{fig_8}(b). This indicates that the reverse drift velocity of electrons suppresses the PSHE shift of the reflected light. For the GH shifts at non-phase-jump points, the magnitudes for both $S$- and $P$-polarized light decrease with increasing $v_D$. In contrast, at phase-jump points, the magnitude of the GH shift remains almost unchanged, though its position shifts slightly, as seen in panels (c) and (d) of FIG.~\ref{fig_8}.

Next, we investigate the wavelength dependence of light shifts under electron reverse drift ($v_D= -0.35v_F$). FIG.~\ref{fig_9} illustrates the wavelength-dependent reflection coefficients and light shifts of the reflected beam in reverse drift mode. As shown in FIG.~\ref{fig_9}(a), the reflection coefficients for both $S$- and $P$-polarized light are modulated by both $\lambda$ and $\theta_i$. 
When $\lambda$ increases, the separation between the primary and secondary peaks of the PSHE shift for reflected $P$-polarized light in the region beyond the Brewster angle significantly widens. With increasing wavelength $\lambda$, the PSHE shift for reflected P-polarized light demonstrates pronounced broadening of the primary-secondary peak separation in the post-Brewster-angle regime. For $\lambda=650$ nm, the minimum reflection coefficient of $P$-polarized light near the Brewster angle is flattened, resulting in a notable reduction of the primary PSHE shift peak at this angle. Concurrently, the positions of the secondary peaks shift to the right as $\lambda$ increases. Additionally, a significantly enhanced negative PSHE shift is observed near $\theta=50^\circ$ for $\lambda=650$ nm. This phenomenon arises from the structural resonance at varying wavelengths and incident angles.

The increase in wavelength also induces pronounced changes in the large GH shifts for both $S$- and $P$-polarized light. For $S$-polarized light (FIG.~\ref{fig_9}(d)): The number of large GH shift positions decreases from 7 (at $\lambda=550$ nm) to 6 ($\lambda=600$ nm) and then to 4 ($\lambda=650$ nm). At $\lambda=600$ nm, the magnitude of negative large GH shifts is strongly suppressed, reaching only 1/10 of those observed at $\lambda=550$ nm and $\lambda=600$ nm. The positions of large GH shift first shift leftward and then rightward as $\lambda$ increases. For P-polarized light (FIG.~\ref{fig_9}(e)): The number of large GH shift positions increases with $\lambda$, opposite to the trend for $S$-polarized light. A negative large GH shift emerges at $\lambda=650$ nm.
\subsection{\label{sec:level2}PSHE and GH shift with Co-directional Fizeau drag}
\begin{figure}
	\centering
	\includegraphics[width=1\linewidth]{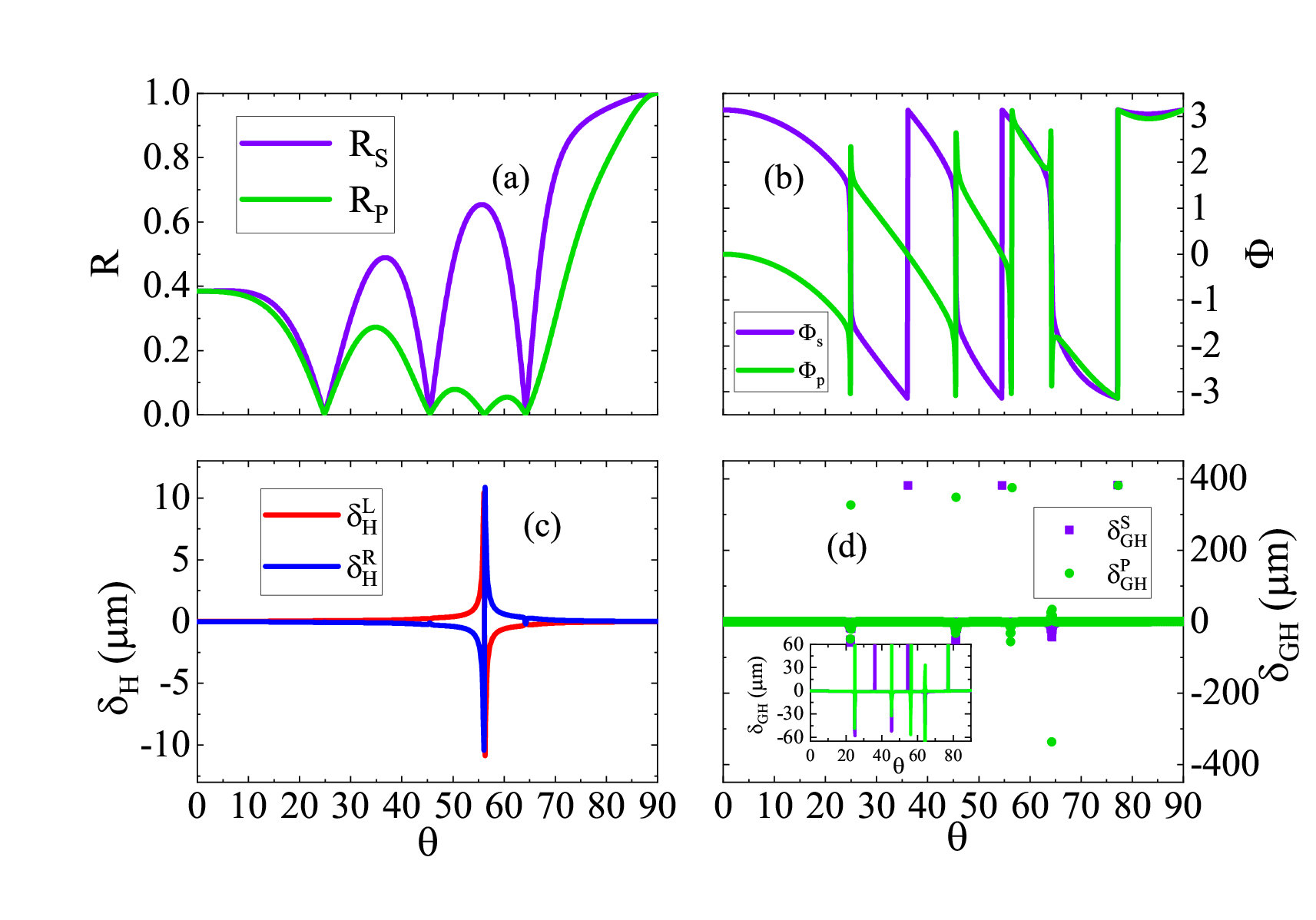}
	\caption{
		Dependence of the (a) Reflection coefficients for $S$- and $P$-polarized light, (b) PSHE shift for $P$-polarized light, (c) GH shift for $S$-polarized light, and (d) GH shift for $P$-polarized light on the incident angle $\theta_i$ with co-directional Fizeau drag ($v_D = 0.35 v_F$). The purple and green solid lines correspond to $S$- and $P$-polarized light, while the red and blue solid lines correspond to left- and right-circularly polarized light. The parameters are set as $\lambda = 600$ nm, $\varepsilon_3 = 2.25$, and $\varepsilon = \varepsilon_{yy}$.	
	}
	\label{fig_10}
\end{figure}

Next, we consider the case where the drift direction of Dirac electrons in the semi-Dirac material aligns with the light incidence direction at $v_D=0.35v_F$. FIG.~\ref{fig_10}(a) demonstrates a significant reduction in the minimum reflection coefficient, which is consistent with the results obtained in the reverse mode. Notably, the co-directional mode exhibits a more substantial decrease in reflectance. As shown in FIG.~\ref{fig_10}(c), the PSHE shift for $P$-polarized light increases, with large shifts localized near the Brewster angle. The magnitude of these shifts exceeds those observed under reverse drift conditions. In the co-directional drift mode, the number of phase-jump points for P-polarized light increases, leading to a greater number of large GH shifts, including negative shifts. Notably, at these newly emerged phase-jump points, the magnitude of the GH shift for $P$-polarized light is always smaller than that for $S$-polarized light. This phenomenon indicates that co-directional drift introduces additional phase-jump points, resulting in diverse and complex GH shift behaviors. For $S$-polarized light, the negative phase-jump points originally present in FIG.~\ref{fig_6}(b) disappear under co-directional drift. Consequently, only three positive large GH shifts remain for $S$-polarized light, while its negative GH shifts are strongly suppressed.

\begin{figure}
	\centering
	\includegraphics[width=1\linewidth]{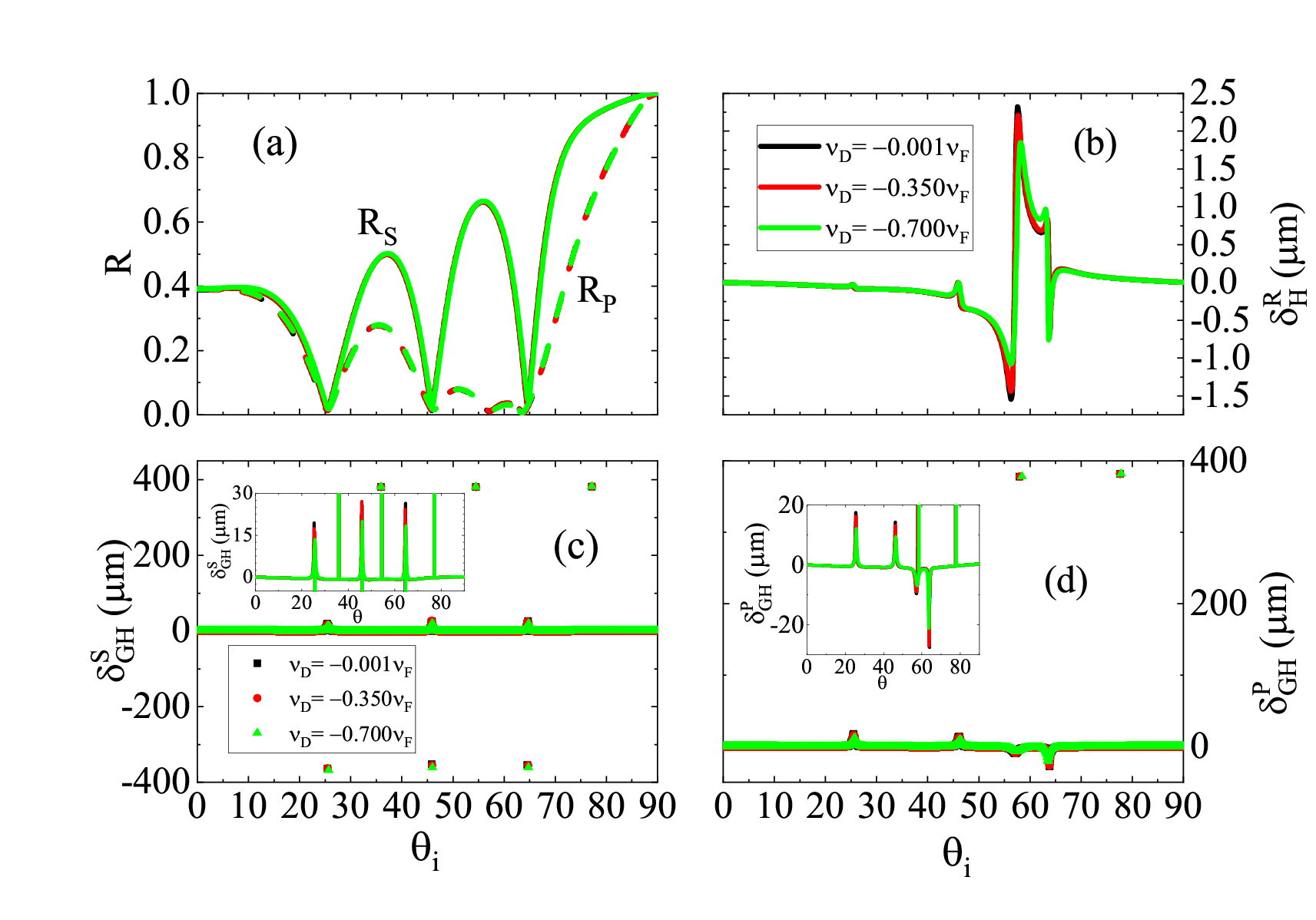}
	\caption{
		 Dependence of (a) Reflection coefficients for $S$- and $P$-polarized light, (b) PSHE shift for $P$-polarized light, (c) GH shift for $S$-polarized light, and (d) GH shift for $P$-polarized light on the incident angle $\theta_i$ with different reverse Fizeau drag. The insets in (c) and (d) display the GH shifts for $S$- and $P$-polarized light at non-phase-jump points. In the figure, black, red, and green lines (or dots) represent cases for drift velocities of $v_D=-0.001,-0.35,-0.7v_F$, respectively.		
	}
	\label{fig_8}
\end{figure}

\begin{figure}
	\centering
	\includegraphics[width=1\linewidth]{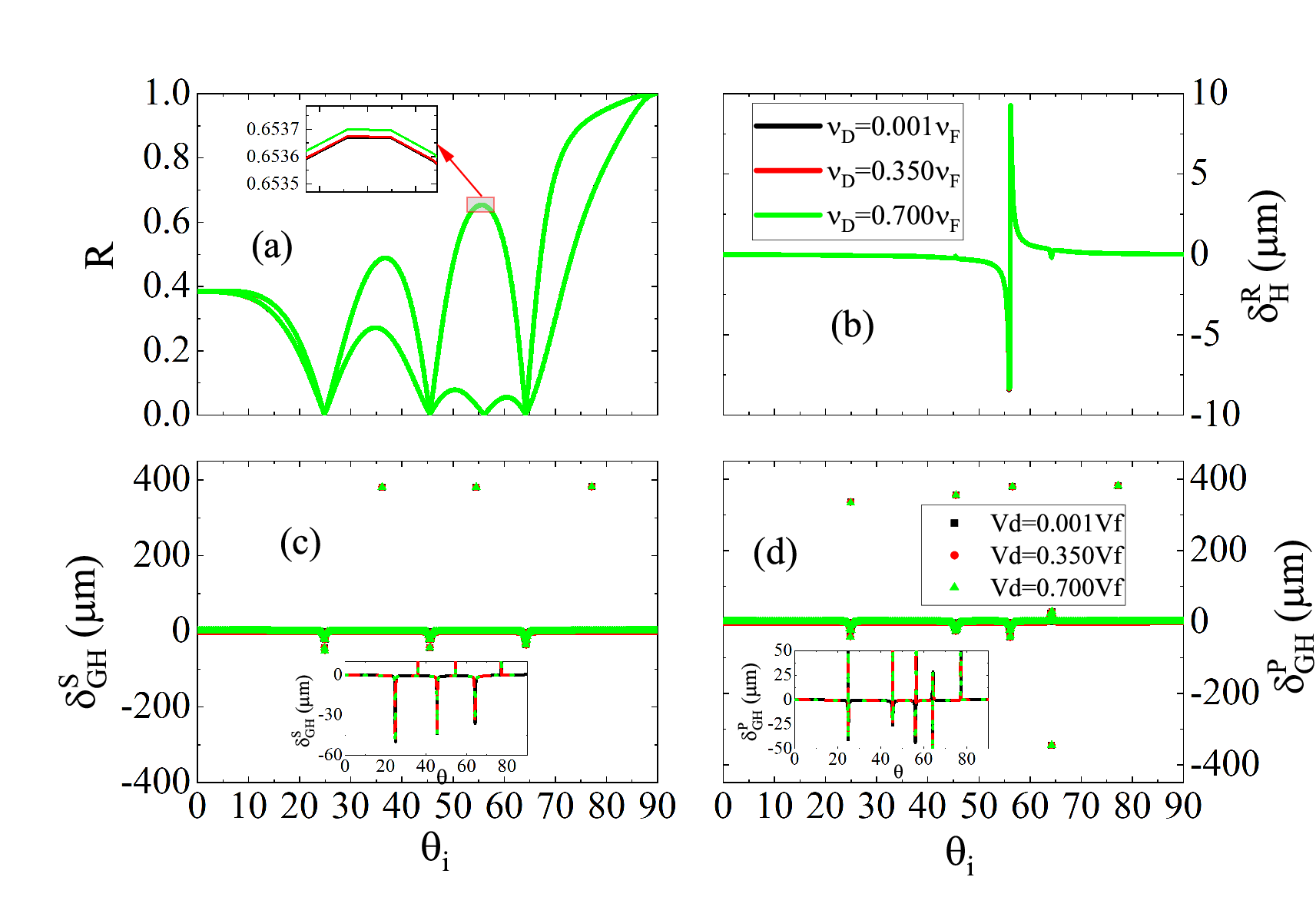}
	\caption{
		Dependence of the (a) Reflection coefficients for $S$- and $P$-polarized light, (b) PSHE shift for $P$-polarized light,FIG.~\ref{fig_13}(a) is a zoom-in view near the Brewster angle. (c) GH shift for $S$-polarized light, and (d) GH shift for $P$-polarized light on the incident angle $\theta_i$ with different co-directional Fizeau drag. The insets in (c) and (d) display the GH shifts for $S$- and $P$-polarized light at non-phase-jump points. In the figure, black, red, and green lines (or dots) represent cases for drift velocities of $v_D = 0.001, 0.35, 0.7 v_F$, respectively.	
	}
	\label{fig_11}
\end{figure}

Under co-directional drift, increasing the drift velocity $v_D$ yields the results shown in FIG.~\ref{fig_11}. The reflection coefficients for $S$- and $P$-polarized light remain nearly unchanged with higher $v_D$, while their corresponding optical shifts slightly decrease. By comparing these results with FIG.~\ref{fig_8}, we conclude that the influence of $v_D$ on the system in co-directional drift mode is significantly weaker than in reverse drift mode.

\begin{figure}
	\centering
	\includegraphics[width=1\linewidth]{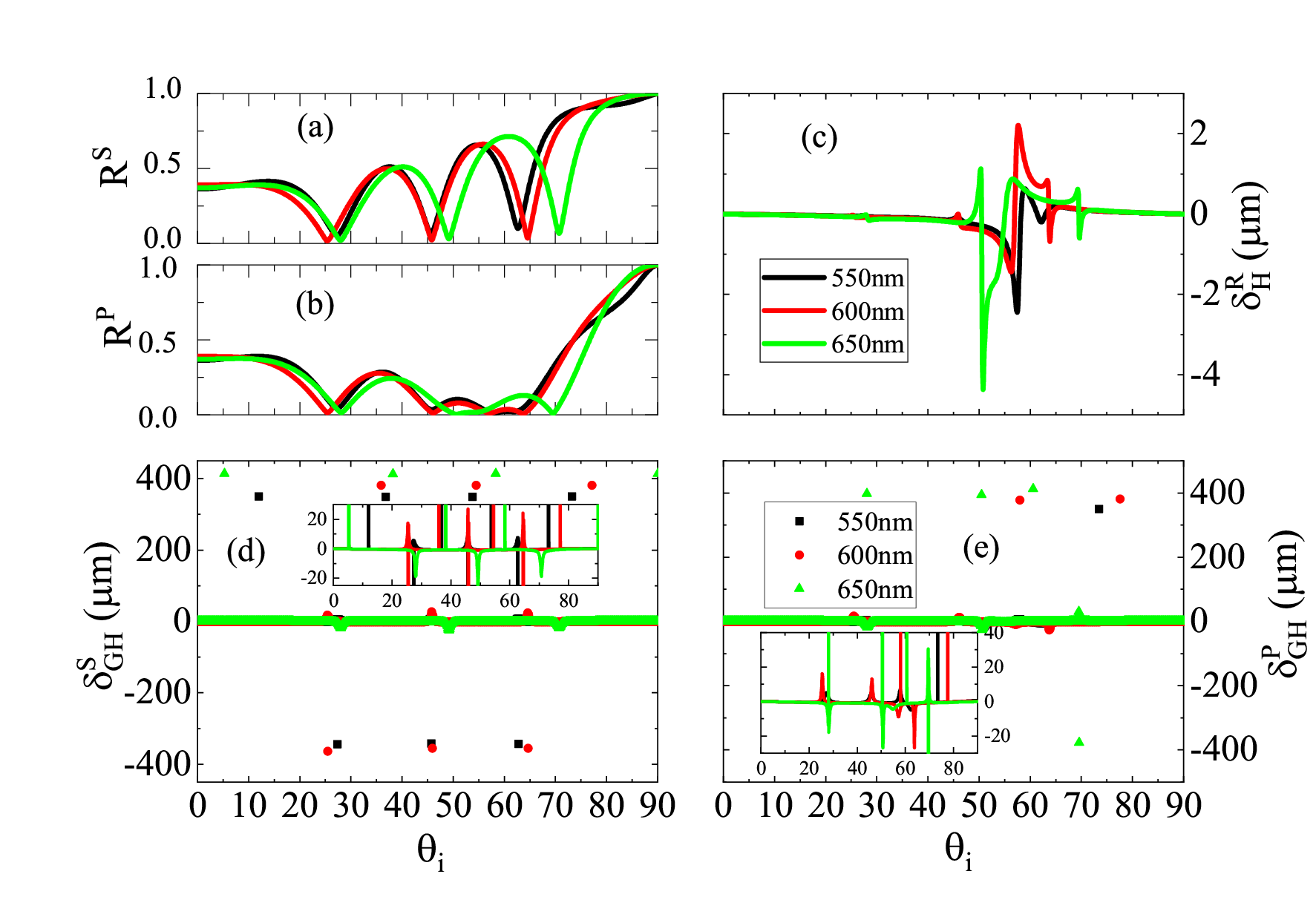}
	\caption{
			Angular dependencies under reverse Fizeau drag ($v_D = -0.35 v_F$): (a,b) Reflection coefficients for $S$- and $P$-polarized light, (c) photonic spin Hall effect (PSHE) shift of $P$-polarization, (d,e) Goos-Hänchen (GH) shifts for (d) $S$- and (e) $P$-polarized states. Insets in (d,e) show GH shifts at non-phase-jump regimes. The lines (or dots) in different colors correspond to wavelengths $\lambda = 550, 600, 650$ nm.		
	}
	\label{fig_9}
\end{figure}

\begin{figure}
	\centering
	\includegraphics[width=1\linewidth]{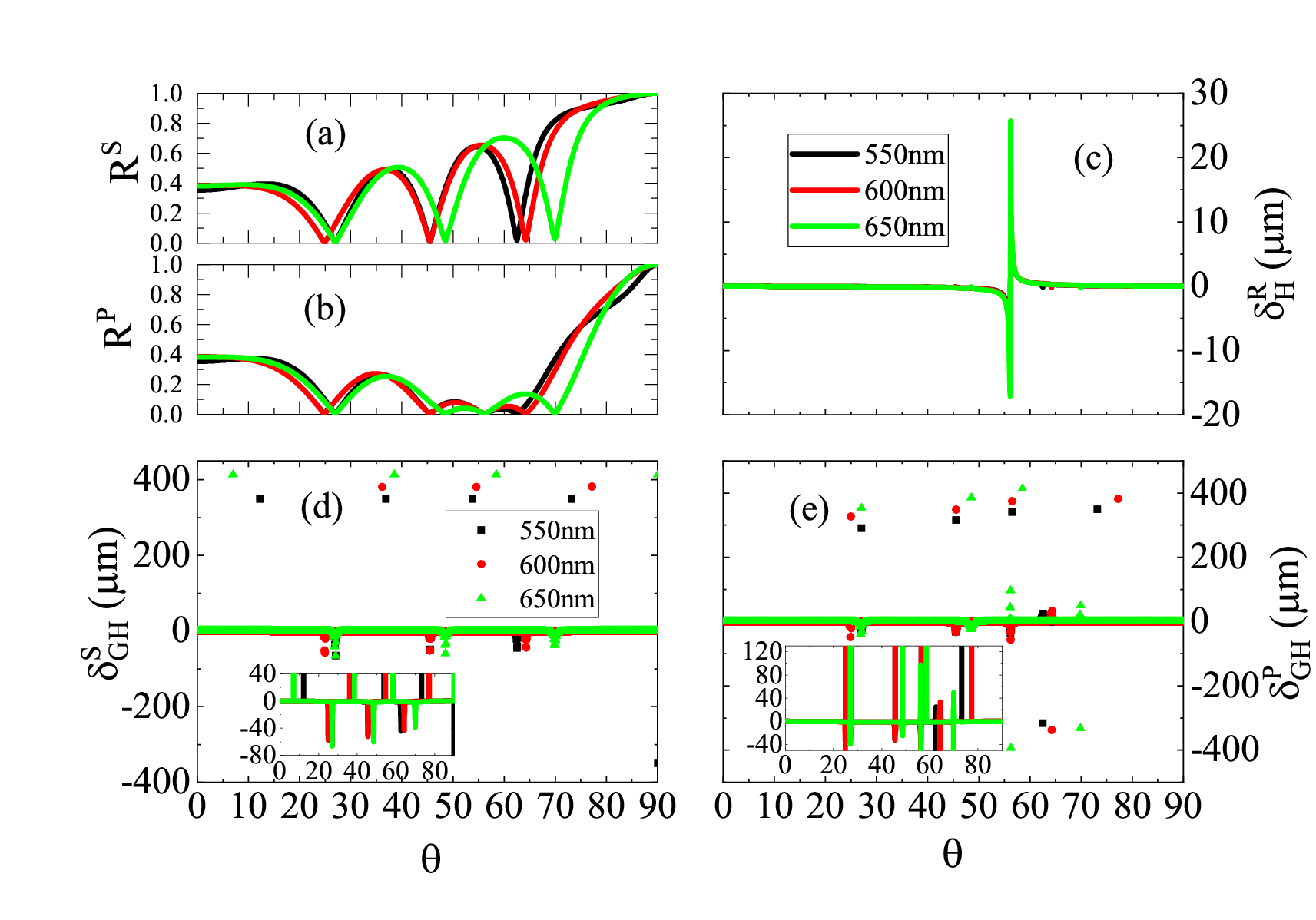}
	\caption{
		Angular dependencies under co-directional Fizeau drag ($v_D = 0.35 v_F$): (a,b) Reflection coefficients for $S$- and $P$-polarized light, (c) photonic spin Hall effect (PSHE) shift of $P$-polarization, FIG.~\ref{fig_13}(b) is a zoom-in view near the Brewster angle.(d,e) Goos-Hänchen (GH) shifts for (d) $S$- and (e) $P$-polarized states. Insets in (d,e) show GH shifts at non-phase-jump regimes. The lines (or dots) in different colors correspond to wavelengths $\lambda = 550, 600, 650$ nm.	
	}
	\label{fig_12}
\end{figure}

FIG.~\ref{fig_12} illustrates the effects on the reflected beam in co-directional drift mode. As shown in panels (a) and (b), the variations in reflection coefficients for $S$- and $P$-polarized light in co-directional drift mode align with those in reverse drift mode, being modulated by both $\lambda$ and $\theta_i$. Near the Brewster angle, the PSHE shift is significantly enhanced, with its magnitude increasing as $\lambda$ grows and exhibiting large amplitude variations. For large GH shifts, the positions and magnitudes of the shifts for $S$- and $P$-polarized light change with $\lambda$. At $\lambda=650$ nm, the large GH shifts for both polarizations increase in magnitude and shift rightward, consistent with the results observed in reverse drift mode (FIG.~\ref{fig_9}). Notably, negative large GH shifts for $S$-polarized light completely vanish under co-directional drift. At shorter wavelengths (e.g. $\lambda=550$ nm), the GH shifts for $P$-polarized light in co-directional drift display more complex features, including additional peaks and enhanced sensitivity to $\theta_i$.

\begin{figure}
	\centering
	\includegraphics[width=1\linewidth]{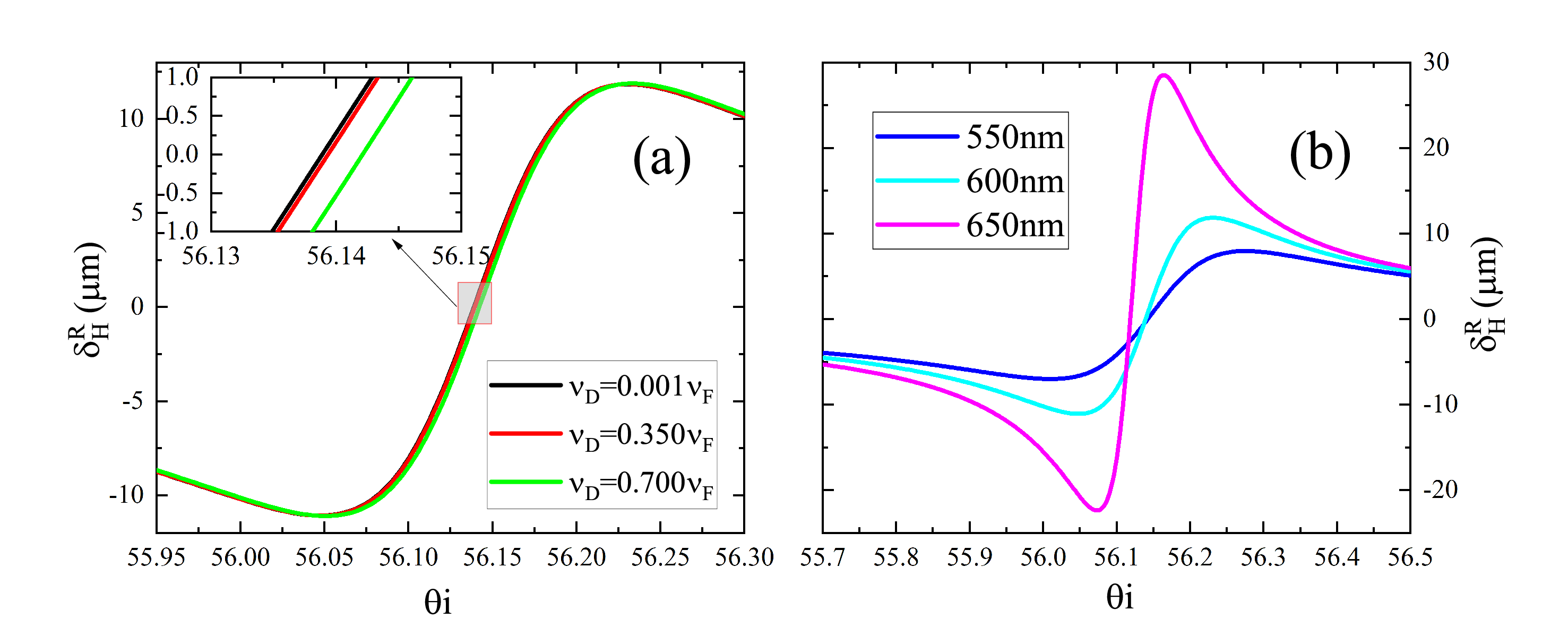}
	\caption{The enlarged views near the Brewster angle from FIG.~\ref{fig_11} and FIG.~\ref{fig_12}.}
	\label{fig_13}
\end{figure}
\section{\label{sec:level1}	conclusion}
This study elucidates the modulation mechanisms of semi-Dirac materials on both the PSHE and GH shifts. Through theoretical investigations, we demonstrate the critical role of the anisotropic dielectric function in semi-Dirac materials in controlling the beam displacements of reflected $S$- and $P$-polarized light. The research shows that the combined effects of incident light wavelength ($\lambda$) and angle ($\theta_i$) induce nearly periodic variations in reflection coefficients and their phases. Furthermore, by leveraging the Fizeau drag effect induced by unidirectional drift of massless Dirac electrons in semi-Dirac materials, we achieve dynamic tuning of optical shifts through the magnitude and direction of drift velocity. Results indicate that in reverse drift mode, increasing the electron drift velocity significantly suppresses the PSHE shift for $P$-polarized light, while GH shifts for both $S$- and $P$-polarized light exhibit minimal changes. By contrast, in co-directional drift mode, the PSHE shift is markedly enhanced, with pronounced changes in beam displacements and phase-jump points near the Brewster angle. The unique electronic properties of semi-Dirac materials—particularly their electron drift effects and anisotropic dielectric response—provide novel pathways for manipulating optical displacements and polarization-dependent phenomena.

\begin{acknowledgments}
The work is supported by Basic Research fund of Nanjing University of Aeronautics and Astronautics with grant No.56XCA2405003. Hong-Liang Cheng acknowledges support from Postgraduate Innovation Program of Nanjing University of Aeronautics and Astronautics with grant No. xcxjh20242119.
\end{acknowledgments}

\appendix*

\section{Derivation of the wave vector in semi-Dirac materials under the Fizeau drag effect}

The phase velocity of light influenced by drifting electrons in semi-Dirac materials is given by: 
\begin{equation}\label{eqA1}
v^m = v \pm v_D F
\end{equation}

 $v$ is defined as: $v = \frac{\omega}{k}$. $\omega^m$ and $k_{SD}^m$ are the angular frequency and wave vector after Lorentz transformation, they can be expressed as $\omega^m = \gamma \left(\omega - v_D k_2 \right)$ and $k_{SD}^m = \gamma \left(k_2 - \frac{v_D}{v_F^2} \omega \right)$. Thus, the phase velocity becomes: $\frac{\omega^m}{k_{SD}^m} = \frac{\omega}{k} \pm v_D F$.
 
Eq.(\ref{eqA1}) can be written as:
\begin{equation}
\frac{A}{k_{SD}^m} - \frac{B}{C k_{SD}^m + D} \mp E = 0
\end{equation}

Here, $A = \gamma \left[ \omega - v_D k_{SD}^m (w) \right]$, $B = \gamma \omega c^2$, $C = c^2$, $D = \gamma \omega v_D$, $E = v_D F(\omega)$.

First consider the co-propagating:
\begin{equation}
\frac{A}{k_{SD}^+} - \frac{B}{C k_{SD}^+ + D} - E = 0
\end{equation}

Expanding the terms:
\begin{equation}
CE {k_{SD}^+}^2 + (DE + B - AC) k_{SD}^+ - AD = 0
\end{equation}

Thus, the solution for \(k_{SD}^+\) is:
\begin{equation}
	\begin{split}
k_{SD}^+ = \frac{1}{2CE}\bigg[
(-(DE + B - AC) \pm  \\
\sqrt{
	\begin{aligned}[t]
		(DE + B - AC)^2 - 4CE (-AD)
	\end{aligned}
} \bigg]
	\end{split}
\end{equation}

\subsubsection*{Part One: The first term of the numerator}

We have the expression:
\begin{equation}
DE + B - AC = \gamma \omega v_D^2 F + \gamma v_D k_{2} c^2
\end{equation}
\subsubsection*{Part Two: The expression under the radical}

\begin{equation}
	(DE + B - AC)^2 - 4CE (-AD)
\end{equation}

Now expand \((DE + B - AC)^2\):
\begin{equation}
	\begin{split}	
(DE + B - AC)^2 = \left( \gamma \omega v_D^2 F + \gamma v_D k_{2} c^2 \right)^2\\
= \gamma^2 \omega^2 v_D^4 F^2 + \gamma^2 v_D^2 k_{2}^2 c^4 + 2 \gamma^2 v_D^3 c^2 \omega F k_{2}
 \end{split}
\end{equation}

The term \(4CE (-AD)\) becomes:
\begin{equation}
	\begin{split}	
4CE (-AD) = 4c^2 v_D F (- \gamma [\omega - v_D k_{2}] \gamma w v_D)\\
= -4 \gamma^2 v_D^2 c^2 \omega^2 F + 4 \gamma^2 c^2 v_D^3 \omega F k_{2}
\end{split}
\end{equation}

Thus, we have
%\[(DE + B - AC)^2 - 4CE (-AD) \]
%\begin{equation}
%	\begin{split}
%= \gamma^2 \omega^2 v_D^4 F^2 + \gamma^2 v_D^2 k_{2}^2 c^4 \\
%- 2 \gamma^2 v_D^3 c^2 \omega F k_{2} + 4 \gamma^2 v_D^2 c^2 \omega^2 F
%	\end{split}
%\end{equation}
\begin{equation}
	\begin{split}
(DE + B - AC)^2 - 4CE (-AD) = \gamma^2 \omega^2 v_D^4 F^2 \\
+ \gamma^2 v_D^2 k_{2}^2 c^4 
	- 2 \gamma^2 v_D^3 c^2 \omega F k_{2} + 4 \gamma^2 v_D^2 c^2 \omega^2 F
		\end{split}
\end{equation}

\subsubsection*{Part three: The denominator term}
\begin{equation}
2CE = 2 c^2 v_D F = 2 c^2 v_D F
\end{equation}

Final expression for \(k_{SD}^+\):
%\begin{widetext}
%	\begin{align}
%k_{SD}^+ = \frac{
%	- (v_D \gamma \omega F + \gamma v_F^2 k_{2})v_D 
%	\pm \gamma v_D \sqrt{
%		k_{2}^2 v_F^4 
%		- 2 v_F^2 v_D \omega k_{2} F 
%		+ 4 v_F^2 \omega^2 F 
%		+ v_D^2 \omega^2 F^2
%	}
%}{
%	2 v_F^2 v_D F
%} 
%	\end{align}
%\end{widetext}
\begin{equation}
	\begin{split}
		k_{SD}^+ = \frac{1}{2 v_F^2 v_D F} \bigg[
		(-v_D \gamma \omega F - \gamma v_F^2 k_{2})v_D\pm \gamma v_D  \\
		\sqrt{
			\begin{aligned}[t]
				k_2^2 v_F^4 - 2 v_F^2 v_D \omega k_2 F + 4 v_F^2 \omega^2 F + v_D^2 \omega^2 F^2
			\end{aligned}
		} \bigg]
	\end{split}
\end{equation}

This is the wave vector in semi-Dirac material in co-propagating mode. Through identical operations, one obtains the wave vector in semi-Dirac material in counter-propagating mode:
%\begin{widetext}
%	\begin{align}
%	k_{\mathrm{SD}}^- &= \frac{
%	(-v_D \gamma \omega F + \gamma v_F^2 k_{2})v_D 
%	\pm \gamma v_D \sqrt{
%		k_{2}^2 v_F^4 
%		+ 2 v_F^2 v_D \omega k_{2} F 
%		- 4 v_F^2 \omega^2 F 
%		+ v_D^2 \omega^2 F^2
%	}
%}{
%	2 v_F^2 v_D F
%}
%	\end{align}
%\end{widetext}
\begin{equation}
	\begin{split}
		k_{SD}^- = \frac{1}{2 v_F^2 v_D F} \bigg[
		(-v_D \gamma \omega F + \gamma v_F^2 k_{2})v_D\pm \gamma v_D  \\
		\sqrt{
			\begin{aligned}[t]
				k_2^2 v_F^4 + 2 v_F^2 v_D \omega k_2 F - 4 v_F^2 \omega^2 F + v_D^2 \omega^2 F^2
			\end{aligned}
		} \bigg]
	\end{split}
\end{equation}\\
\bibliographystyle{unsrt}
\bibliography{apstemplate}
\end{document}